\documentclass{article}
\usepackage{spconf,amsmath,graphicx}
\usepackage{multirow}
\usepackage{multicol}
\usepackage{hyperref}
\usepackage{todonotes}
\usepackage{xcolor,colortbl}
\usepackage{float}
\usepackage{booktabs}
\usepackage{authblk}
\usepackage{stfloats}

\newcolumntype{L}[1]{>{\raggedright\let\newline\\\arraybackslash\hspace{0pt}}m{#1}!{\hspace{3pt}}}
\newcolumntype{C}{>{\color{white}\columncolor{gray!30}[.5\tabcolsep]%
\centering\arraybackslash}p{20mm}}

\title{An Investigation of End-to-End Models for Robust Speech Recognition}
\name{Archiki Prasad$^\dagger$, Preethi Jyothi$^\ddagger$, and Rajbabu Velmurugan$^\dagger$}
  
\address{$^\dagger$ Department of Electrical Engineering, Indian Institute of Technology Bombay \\ $^\ddagger$ Department of Computer Science \& Engineering, Indian Institute of Technology Bombay}

\begin{document}

\begin{minipage}{0.9\textwidth}
{{\textcopyright} 2021 IEEE. Personal use of this material is permitted. Permission from IEEE must be obtained for all other uses, in any current or future media, including reprinting/republishing this material for advertising or promotional purposes, creating new collective works, for resale or redistribution to servers or lists, or reuse of any copyrighted component of this work in other works.} 
\end{minipage}
\clearpage

\maketitle
\begin{abstract}
End-to-end models for robust automatic speech recognition (ASR) have not been sufficiently well-explored in prior work. With end-to-end models, one could choose to preprocess the input speech using speech enhancement techniques and train the model using enhanced speech. Another alternative is to pass the noisy speech as input and modify the model architecture to adapt to noisy speech. A systematic comparison of these two approaches for end-to-end robust ASR has not been attempted before. We address this gap and present a detailed comparison of speech enhancement-based techniques and three different model-based adaptation techniques covering data augmentation, multi-task learning, and adversarial learning for robust ASR. While adversarial learning is the best-performing technique on certain noise types, it comes at the cost of degrading clean speech WER. On other relatively stationary noise types, a new speech enhancement technique outperformed all the model-based adaptation techniques. This suggests that knowledge of the underlying noise type can meaningfully inform the choice of adaptation technique.
\end{abstract}
%

\section{Introduction}
\label{sec:intro}

End-to-end (E2E) models, that directly convert a spoken utterance into a sequence of characters, are becoming an increasingly popular choice for ASR systems. They have been shown to outperform traditional cascaded ASR systems when large amounts of labeled speech are available. E2E ASR systems for low-resource scenarios have also emerged as an active new area of research. While E2E ASR systems for clean speech are growing rapidly in number, there have been relatively  fewer investigations on the use of E2E models for noisy speech recognition~\cite{liang2018learning,zhang2020learning}. To the best of our knowledge, we are the first to provide a detailed comparison of speech enhancement-based techniques with a number of E2E model-based adaptation techniques for noisy speech across a diverse range of noise types. This comparison highlights the strengths and limitations of both types of approaches and offers prescriptions for which techniques are best suited for different noise types. Our code and datasets are publicly available.\footnote{\url{https://github.com/archiki/Robust-E2E-ASR}}

Prior work on robust ASR has predominantly used a two-pass approach to tackle the problem of robust ASR. The input speech is first passed through a speech enhancement (SE) module and the enhanced speech is subsequently passed through a standard speech recognition system. We adopt this as one of our approaches as well and investigate the use of three different speech enhancement techniques in conjunction with an E2E ASR system. As opposed to modifying the input speech with front-end processing modules like speech enhancement, one could use the noisy speech as-is and adapt the E2E model itself to handle the noisy input speech. We examine three model-based adaptation techniques:

\begin{enumerate}
\item Data Augmentation-based Training (DAT): Speech samples are augmented with varying noise types of varying signal to noise ratio (SNR) values and fed as input to an E2E model. Larger gradient updates are made in the lower layers of the E2E model compared to the higher layers. This technique first appeared in~\cite{zhang2020learning}.

\item Multi-task learning (MTL): The E2E model is jointly trained with an auxiliary noise type classifier. MTL with a noise type classifier has not been previously explored for robust ASR and  turns out to be quite effective as an adaptation technique.

\item Adversarial Training (AvT): Unlike MTL which drives the learned representations to be more noise-aware, in AvT, we train an E2E model with a gradient reversal layer in conjunction with the noise classifier to learn more noise-invariant representations.
\end{enumerate}

    \section{Related Work}
\label{sec:rel}
We review some relevant supervised robust ASR techniques under noisy conditions and single-channel setting, and do not consider reverberation or multi-channel approaches. A summary of various deep learning techniques for robust ASR, datasets, and benchmarks are provided in~\cite{zhang2018deep}. 
In~\cite{seltzer2013investigation}, a noise aware training (NAT) technique that uses a mean noise estimate (assuming stationarity) in the input has been proposed to give good results. This has been further improved in~\cite{narayanan2014joint}, by jointly training a source separation model for noise estimation and acoustic modeling.
The work in~\cite{qian2016very} uses deep convolutional neural network (CNN) to achieve the best ASR results on Aurora-4 task~\cite{pearce2002aurora}.

One approach to improve the performance of E2E models is by using data augmentation along with fine tuning as in~\cite{ko2017study}. Another approach is to formulate this as a domain adaptation problem, and then use variational auto encoders (VAE)  ~\cite{hsu2017unsupervised}. Here, the source domain is clean speech and target domain is noisy speech. 
Similarly, \cite{liang2018learning} uses penalty terms on an encoder-decoder type ASR model to achieve noise invariant representations. The work in~\cite{liu2019jointly} uses a jointly adversarial training framework, with a mask-based SE front-end along with an attention-based E2E ASR model.

 Models that operate in time-domain, such as SE-GAN~\cite{pascual2017segan} and SE-VCAE~\cite{braithwaite2019speech} have been effective for SE as a front-end technique in a two-pass approach. However, such models rely on large datasets for training and enhancement, but need not necessarily improve ASR. Deep Xi~\cite{nicolson2019deep}, a recent front-end SE has been shown to provide improved ASR when used with DeepSpeech2~\cite{amodei2016deep} (DS2) as the back-end ASR. A more recent approach that operates on raw waveform for real-time speech enhancement is~\cite{defossez2020real}. These two methods will be further explained in Section~\ref{ssec:SE}, and will be used in our analysis. 




\section{Datasets and E2E ASR System}
\label{sysdata}

\subsection{Implementation Details}
In this work, we use DS2 as our main E2E ASR system. DS2 is trained using the Connectionist Temporal Classification (CTC) objective function and comprises two 2D convolutional layers, followed by five bidirectional LSTM layers with a final fully-connected (FC) softmax layer. The input features are derived from short-time Fourier transform (STFT) magnitude spectrograms and the baseline model is trained on 100 hours of clean speech from the Librispeech Dataset~\cite{panayotov2015librispeech}. Additional training-specific details can be found in ~\cite{prasad2020accents}.\footnote{We use the DS2 implementation available at: \url{https://github.com/SeanNaren/deepspeech.pytorch}}

\subsection{Data Description}
We are specifically interested in applications that cater to certain noise types for which we do not have a lot of data. There are not many existing datasets that are designed to be low-resource.%
\footnote{The DEMAND dataset~\cite{thiemann2013demand} exists but has few samples per noise type.}
Hence, we constructed our own custom dataset.


Our custom dataset consists of the following noise types: \textit{`Babble', `Airport/Station', `Car', `Metro', `Cafe', `Traffic', `AC/Vacuum'}\footnote{\textit{Airport/Station} comprises background sounds containing announcements at both airports and stations, while \textit{AC/Vacuum} comprises room sounds with an air conditioner or vacuum cleaner in the background. Within these noise types, the noise sub-types are equally distributed in train and test sets.}. The noise samples are sampled at 16 kHz and were collected from FreeSound~\cite{font2013freesound}. For each noise type, we have 10 and 8 distinct samples in the train and test sets, respectively. The total duration of the noise train and test sets is close to 2 hours. For clean speech, we use the Librispeech corpus~\cite{panayotov2015librispeech}. We train the DS2 models using the \emph{train-clean-100} set and add simulated noise using the training samples of our noise dataset. Our development set is constructed using the \emph{dev-clean} set of Librispeech (duration of 5.4 hours) with training noise samples from our noise dataset. During both training and validation, a noise type was picked randomly and the SNR of the additive noise was chosen randomly from \{0, 5, 10, 15, 20, 25\}~dB.  For testing, we randomly picked 120 files from the \emph{test-clean} set of Librispeech. To create noisy speech, for each utterance and each noise type from our noise database, a random section of the noise is added at SNR levels ranging from [0, 20] dB in increments of 5dB, resulting in a total of 4200 noisy speech test utterances. (This process of constructing noisy test samples was outlined in~\cite{nicolson2019deep}.)

\section{Approaches} 

\subsection{Front-End Speech Enhancement}
\label{ssec:SE}

We experiment with three state-of-the-art speech enhancement techniques detailed below. 

\noindent \textbf{SE-VCAE.} Speech Enhancement Variance Constrained Autoencoder (SE-VCAE)~\cite{braithwaite2019speech} learns the distribution over latent features given noisy data and acts directly on the time-domain. It  outperforms a popular generative modeling SE technique SE-GAN~\cite{pascual2017segan}. We finetune the pretrained SE-VCAE model on noisy speech samples from our dataset. 


\noindent \textbf{DeepXi.} DeepXi was specifically proposed as a front-end to be used with DeepSpeech ~\cite{nicolson2019deep}. This enhancement technique acts on a noisy speech spectrogram and uses a priori SNR estimation to design a mask that is used to produce an estimate of a clean speech spectrogram. DS2 is then fine-tuned for 10 epochs on the enhanced examples from Librispeech while ensuring minimal loss in performance on clean speech data. 

\noindent \textbf{D\textsc{emucs}.}
 \cite{defossez2020real} proposes an alternate encoder-decoder architecture for denoising speech samples. The model is trained on a low-resource noisy dataset along with reverberation, with data augmentation techniques to compensate for the limited data. We use their pretrained model and fine-tune it on our dataset for 20 epochs. Similar to DeepXi, DS2 is further finetuned on the denoised samples. 
\begin{table}[t]
    \centering
    \footnotesize
    \setlength\tabcolsep{8 pt}
    \begin{tabular}{l|cccc}
    \toprule
     \multirow{2}{*}{\textbf{Method} } & \multicolumn{4}{c}{\textbf{Objective Scores}}\\
    \cline{2-5}
    & \bf MOS-LQO & \bf PESQ & \bf STOI & \bf eSTOI \\
    \hline
         Baseline & 1.29 & 1.80 & 80.68 & 60.66 \\
         SE-VCAE & 1.38 & 1.83 & 80.64 & 62.85 \\
         Deep Xi & 1.98 & 2.50 & 86.89 & 74.52 \\
         D\textsc{emucs} & 2.17 & 2.73 & 91.66 & 82.19 \\
    \bottomrule
    \end{tabular}
    \caption{Objective scores for various enhancement methods. Larger scores are better.}
    \label{tab:enhance}
\end{table}
Table~\ref{tab:enhance} shows objective scores measuring the quality of enhanced speech using the three SE techniques discussed here. D\textsc{emucs} clearly outperforms the other two techniques on all four metrics. 


\subsection{Data Augmentation-based Training (DAT)}
\label{sssec:aug-train}

For this technique, clean speech samples are augmented with noise with a probability of 0.5 and subsequently used to train DS2. The model was trained for 25 epochs with a batch size of 32 and a learning rate of 0.0001. In the end, the model that performed the best on the development set (with similar noise augmentation) was chosen. We will refer to this model as ``Vanilla DAT''. 
To enable better transfer learning from clean to noisy speech, we incorporate the soft-freezing scheme proposed in \cite{zhang2020learning}. The learning rate of the FC layer along with the last two LSTM layers is scaled down by factor of 0.5 (further discussed in Section~\ref{sec:results}). This training strategy has the effect of forcing the lower layers (that act as a feature extractor) to learn noise-invariant characteristics in the noisy speech. This model will henceforth be referred to as ``Soft-Freeze DAT''. 

\subsection{Multi-Task Learning (MTL)}
\label{ssec:mtl}
\begin{figure}[t!]
    \centering
    \includegraphics[scale=0.27]{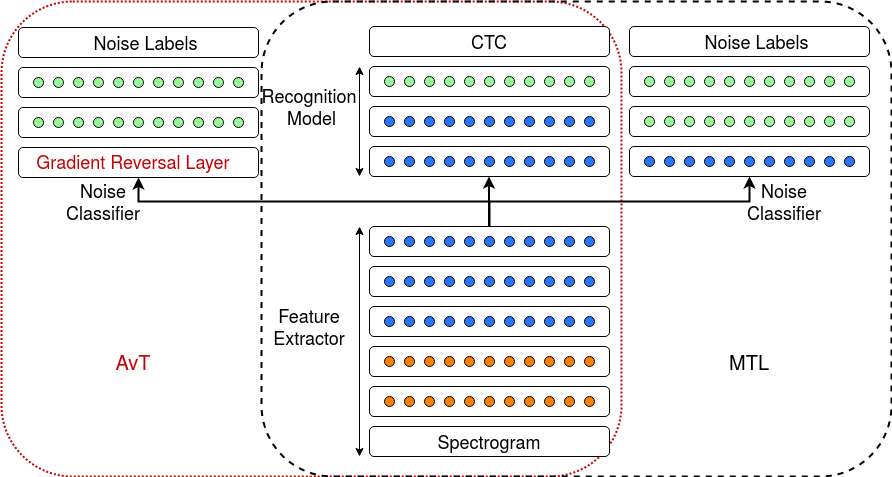}
    \caption{Framework used in MTL (in black) and AvT (in red). 2D-convolutions, BiLSTM and FC layers are shown using orange, blue and green circles, respectively.}
\label{fig:mtl}
\end{figure}
Figure~\ref{fig:mtl} describes our MTL setup. The auxiliary classifier predicts noise type labels\footnote{There are 8 noise labels: 7 for noise types + 1 for clean speech} and uses representations from an intermediate LSTM layer as its input. This noise classifier comprises one bidirectional LSTM layer followed by two linear layers. The model is trained with a hybrid loss, $L_H = \lambda L_{\text{CTC}} + \eta (1 - \lambda) L_{\text{CE}}$, where $L_{\text{CTC}}$ and $L_{\text{CE}}$ are the CTC loss from DS2 and the cross entropy loss on the noise labels, respectively. $\eta$ and $\lambda$ are scaling factors and $\eta$ is annealed by factor of 1.05 every epoch. In our experiments, we initialized the model using Soft-Freeze DAT\footnote{We observed that on starting with the baseline model, the initial 10-15 epochs behaved similar to DAT. Hence, Soft-Freeze DAT served as a good initialization and led to faster convergence.} and set $\lambda = 0.7$, $\eta=10$.%

%

\noindent

\begin{table*}[t!]
    \footnotesize
    \centering
    \begin{minipage}{\linewidth}
    \setlength\tabcolsep{3 pt}
    \begin{tabular}{l|ccccc|ccccc|ccccc|ccccc}
    
    \toprule
           \multirow{3}{*}{\textbf{Method} } & \multicolumn{20}{c}{\textbf{WER under SNR (dB)}}\\
    \cline{2-21}
    & \multicolumn{5}{c|}{\textbf{Babble}} & \multicolumn{5}{c|}{\textbf{Airport/Station}} & \multicolumn{5}{c|}{\textbf{AC/Vacuum}} & \multicolumn{5}{c}{\textbf{Cafe}}\\
    \cline{2-21}
             &      \textbf{0}     & \textbf{5}  &   \textbf{10} &  \textbf{15}   & \textbf{20} &      \textbf{0}     & \textbf{5}  &   \textbf{10} &  \textbf{15}   & \textbf{20} &      \textbf{0}     & \textbf{5}  &   \textbf{10} &  \textbf{15}   & \textbf{20} &      \textbf{0}     & \textbf{5}  &   \textbf{10} &  \textbf{15}   & \textbf{20}\\
         \hline
Baseline & 104.2 & 98.3 & 91.3 & 79.7 & 65.0 & 91.9 & 84.1 	& 73.7 & 60.6 & 50.0 & 93.0 & 	83.1 &71.5 & 59.5 &	45.8& 83.8 & 72.7 & 59.5 & 44.3 & 33.4  \\ \midrule
SE-VCAE&     85.6 &     76.4     &  61.9  & 54.7 & 39.7  &  78.0 & 68.3  & 56.8 &  46.3 &  39.3 & 81.3 & 71.1 &  61.3 &  53.6 & 42.7& 61.6 & 53.9 & 44.9 & 35.8 & 31.0 \\
Deep Xi & 81.4 &  69.4 &  54.0 &  44.5 &  31.9 & 71.4 & 60.9  & 46.5  & 37.8   & 27.4 &  73.9 & 58.2  & 45.4  & 35.1  & 27.0 &  52.3 & 39.7  & 32.8  & 25.0  & 20.4  \\ 
D\textsc{emucs} & \bf 70.3 & \bf 58.0 &  \bf 41.8 & \bf 32.3 & \bf 25.4 & \bf 58.6 & \bf 45.5 & \bf 33.7 & \bf 25.6 & \bf 21.5 & \bf 60.5 & \bf 45.4 & \bf 34.2 & \bf 28.1 & \bf 22.8& \bf 38.9 & \cellcolor{green!20} \bf 31.6 & \bf 27.4 & \cellcolor{green!20}\bf 20.3 & \bf 16.9 \\ \midrule
Vanilla DAT & 80.6 & 68.1 & 53.6 & 41.8 & 30.3 & 67.1 &  55.4 & 41.9 & 31.2 & 24.9 & 66.4 & 49.8 & 38.3 & 31.3 & 24.5& 52.8 & 41.6 & 34.5 & 24.5 & 19.2 \\
Soft-Freeze DAT & 77.4 & 65.5 &   52.2 &38.5 & 28.3 &      64.2 &  52.9 & 39.0 & 29.2 & 23.7 & 63.1 & 46.8 & 37.1 & 30.2 & 23.9& 49.1 & 40.1 & 33.1 &  24.4 &  19.0 \\
MTL & 71.4 & 58.8 & 45.9 & 35.5 & 25.8 & 55.7 &  46.8 & 35.3& 26.2& \cellcolor{green!20}\bf 20.7 & 57.0 &  41.0 &  \cellcolor{green!20} \bf 33.0 & \cellcolor{green!20}\bf 26.7 & \cellcolor{green!20} \bf 21.5& 44.9 & 35.5 & \cellcolor{green!20}\bf 24.8 & 23.6 & \cellcolor{green!20}\bf 15.3 \\
AvT & \cellcolor{green!20} \bf 66.8                  &  \cellcolor{green!20}\bf 55.1  &\cellcolor{green!20} \bf  39.5 &  \cellcolor{green!20}\bf 31.1 &  \cellcolor{green!20} \bf 24.6 & \cellcolor{green!20}\cellcolor{green!20}\bf 53.8 & \cellcolor{green!20}\bf 43.3 & \cellcolor{green!20}\bf 33.4 &\cellcolor{green!20} \bf 25.2 & 20.9 & \bf \cellcolor{green!20}56.4 & \cellcolor{green!20}\bf 40.8 & 33.4 & 29.3 & 23.2& \cellcolor{green!20}\bf 37.1 & \bf 32.0 &  26.3 & \bf 21.4  &  18.5 \\
\bottomrule
\end{tabular}
\end{minipage}

\begin{minipage}{\linewidth}
\vspace{0.35cm}
\centering
\setlength\tabcolsep{3 pt}
 \begin{tabular}{l|ccccc|ccccc|ccccc|c}
    \toprule
        \multirow{3}{*}{\textbf{Method} } & \multicolumn{15}{c|}{\textbf{WER under SNR (dB)}} & \multirow{3}{0.7cm}{\textbf{Clean \\ WER}}\\
    \cline{2-16}
    & \multicolumn{5}{c|}{\textbf{Traffic}} & \multicolumn{5}{c|}{\textbf{Metro}} & \multicolumn{5}{c|}{\textbf{Car}} & \\
    \cline{2-16}
             &      \textbf{0}     & \textbf{5}  &   \textbf{10} &  \textbf{15}   & \textbf{20} &      \textbf{0}     & \textbf{5}  &   \textbf{10} &  \textbf{15}   & \textbf{20} &      \textbf{0}     & \textbf{5}  &   \textbf{10} &  \textbf{15}   & \textbf{20} & \\
         \hline
Baseline &72.4 &	62.5 &	50.2 &	41.0 &	33.6 & 68.4 &	54.4 & 	46.4 &34.9 &27.6 & 35.0 & 28.1 &24.3 & 21.7 & 16.7  & 10.3\\ \midrule
SE-VCAE & 60.0 & 51.9 &  44.2 & 39.7 & 32.5  &  54.0 &  43.6 & 38.6 & 33.0 &  29.6 & 35.4 & 32.7 &  28.3 & 27.3 & 26.0& 15.9\\
Deep Xi & 48.0 & 40.6 & 29.8 & 26.0 & 22.7 & 44.8 & 30.5  & 28.1  & 20.2  & 20.5  & 23.0 & 19.4  & 15.4  & 16.0  & 14.1 & 10.9 \\ 
D\textsc{emucs} &\cellcolor{green!20} \bf 38.2 & \cellcolor{green!20}\bf 30.3 &\cellcolor{green!20} \bf 25.3 &\cellcolor{green!20} \bf 20.6 & \cellcolor{green!20}\bf 17.9 & \cellcolor{green!20}\bf 35.6 & \cellcolor{green!20}\bf 24.9 & \cellcolor{green!20} \bf 22.6 &\cellcolor{green!20} \bf 17.1 &\cellcolor{green!20} \bf 15.9  & \cellcolor{green!20}\bf 20.5 & \cellcolor{green!20}\bf 18.1 & \cellcolor{green!20}\bf 14.6 &\cellcolor{green!20} \bf 13.8 &\cellcolor{green!20} \bf 13.1& 10.9\\ \midrule 
Vanilla DAT & 48.5 & 37.8 & 31.2 & 23.3 & 21.8& 41.8  &  33.1& 27.1&   21.9 &  19.1  & 24.0 & 19.4 & \bf 16.1 & 16.4 & \bf 14.0 & 10.8 \\
Soft-Freeze DAT & 47.2 & 35.1 &  29.8 & 23.4 &  20.2&  40.8 &  30.7 & 27.0 &   21.3 &  18.6 &  23.4 &  \bf 18.9 & 16.8 & 15.2 & 14.7& 10.9\\
MTL &  \bf 39.9 & \bf 32.4 & 29.4 &\bf  21.3 &  \bf 18.4  &  38.7 &     29.2 & 24.4 & 20.6  &  \bf 17.3 &  22.9 & 19.0  & 16.3 &  \bf 14.7 & 14.5  & 11.0\\
AvT & 40.7 & 32.5 & \bf 26.3 &   21.4 & 18.5 & \bf 36.1  & \bf 26.5 & \bf 22.6 & \bf 18.4 & 17.8 & \bf 21.8  &  \bf18.9  &  16.8  &  16.0 & 15.3 & 13.1 \\
\bottomrule
    \end{tabular}
    \end{minipage}
    \caption{Comparison of the performance (WER \% after greedy decoding) of all techniques for various noise types and SNRs in the test set. The lowest SE and E2E WERs are shown in bold, and the lower WER among the two is highlighted in green.}
    \label{tab:main}
\end{table*}

\begin{table*}[!b]
    \footnotesize
    \centering
    \begin{minipage}{\linewidth}
    \centering
    \setlength{\tabcolsep}{3pt}
    \begin{tabular}{l|c|cc|cc|cc|cc|cc|cc|cc}
    
    \toprule
           \multirow{3}{*}{\textbf{Method} } & \multirow{3}{0.7cm}{\textbf{Clean \\ WER}} & \multicolumn{14}{c}{\textbf{WER under SNR (dB)}}\\
    \cline{3-16}
    & & \multicolumn{2}{c|}{\textbf{Babble}} & \multicolumn{2}{c|}{\textbf{Airport/Station}} & \multicolumn{2}{c|}{\textbf{Cafe}} & \multicolumn{2}{c|}{\textbf{AC/Vacuum}} & \multicolumn{2}{c|}{\textbf{Car}} & \multicolumn{2}{c|}{\textbf{Metro}}& \multicolumn{2}{c}{\textbf{Traffic}}\\
    \cline{3-16}
             & & \textbf{0} & \textbf{15} & \textbf{0} & \textbf{15} & \textbf{0} & \textbf{15} & \textbf{0} & \textbf{15} & \textbf{0} & \textbf{15} & \textbf{0} & \textbf{15}& \textbf{0} & \textbf{15}      \\
         \hline

 Soft-Freeze DAT (LSTM$_4$) & 11.13 & 79.2 &40.5 &64.7 & 30.6 &  51.6& \bf 24.2& 63.4 & 30.8 & 24.4 & 16.8& 42.2 & 22.8 & 48.1 & 24.5\\
Soft-Freeze DAT (LSTM$_4$ + LSTM$_3$) & \bf 10.90 & \bf 77.4& \bf38.5 & \bf64.2 & \bf29.2 & \bf 49.1& 24.4 & \bf63.1 & 30.2 & 23.4 & \bf15.2 & \bf40.8&\bf21.3 & \bf47.2 & \bf23.4\\
Soft-Freeze DAT (LSTM$_4$ + LSTM$_3$ + LSTM$_2$) & 10.97 & 79.4  & 40.4& 66.3 & 30.8 & 52.0 & 24.3 & 63.4 & \bf29.4 & \bf22.9 & \bf15.2 & 41.5& 22.1 & 47.8 &24.0 \\
\midrule
MTL (noise classifier after LSTM$_1$) & \bf 11.04 & 72.3 &36.5 & 56.2 & 26.5  & 45.9 & \bf 23.3 & 57.9 & 28.2& \bf22.6 & 15.2 & \bf38.5 & \bf20.2 & 44.0 & 21.6\\
MTL (noise classifier after LSTM$_2$) & 11.05 & \bf 71.4 & \bf 35.5 & \bf55.7& \bf26.2 & \bf44.9 & 23.6 & \bf57.0 & \bf26.7 & 22.9 & \bf\bf14.5 & 38.7 & 20.6 &\bf39.9 & \bf21.3 \\
MTL (noise classifier after LSTM$_3$) & 11.24 & 75.8& 38.7& 60.4 &27.5 &49.0 & 23.5 & 60.0 & 28.2 & 23.0 & 15.7 & 40.4 & 21.6 & 46.0 & 22.4 \\
\midrule
&  & \textbf{15} & \textbf{20} & \textbf{15} & \textbf{20} & \textbf{15} & \textbf{20} & \textbf{15} & \textbf{20} & \textbf{15} & \textbf{20} & \textbf{15} & \textbf{20}& \textbf{15} & \textbf{20} \\
\hline
AvT (recognition scaling factor $\lambda_r = 0.2$) & 13.79 & 31.9 & 26.2 & 27.8 & 22.1 & 21.9 &18.9 & 29.1 & 23.5  & 17.1 & 16.2 & 20.7 & 19.4 &  21.9 &   19.2\\
AvT (recognition scaling factor $\lambda_r = 0.1$) & 13.37&\bf 30.8 & 24.9 & 25.4 & 21.7 & 21.7 & \bf 17.8 & \bf 27.9 & 23.3 & 16.2 & 15.6 & 20.3 & \bf 17.1 & \bf 21.8 & 19.0\\
AvT (recognition scaling factor $\lambda_r = 0.05$) & \bf 13.07& 31.1 & \bf24.6 & \bf25.2 & \bf20.9 & \bf21.4 & 18.5 & 29.3 & \bf 23.2 & \bf 16.0 & \bf 15.3 & \bf 18.4 &   17.8 & 22.1 & \bf 18.7\\
\bottomrule
\end{tabular}
\end{minipage}
    \caption{WER \% after greedy decoding under different settings for DAT, MTL and AvT. The best numbers are shown in bold.}
    \label{tab:exp}
\end{table*}

%

\subsection{Adversarial Training (AvT)}
Contrary to MTL where the model jointly minimizes the CTC loss and the noise classification loss, adversarial training invokes the use of a gradient reversal layer (GRL)~\cite{ganin2015unsupervised} before the auxiliary classifier as shown in Fig~\ref{fig:mtl}. This forces the representations before the GRL to be noise-invariant, thus making it hard for the noise classifier to distinguish between noise types\footnote{A setup similar to AvT was explored in \cite{liang2018learning} using the Musan corpus~\cite{snyder2015musan}. Due to lack of noise labels, the classifier output 2 labels: clean and noisy.}. AvT had to be carefully trained with setting different learning rates, $\lambda_f, \lambda_r$ and $\lambda_n$ corresponding to the feature extractor, recognition model and noise classifier, respectively. For our model, the base learning rate was set to $0.0008$, $\lambda_f = 0.8, \lambda_r = 0.05$ and $\lambda_n = 1$. Similar to MTL, we initialized the model with Soft-Freeze DAT. 


\section{Experiments and Results}

\label{sec:results}

Table~\ref{tab:main} lists an exhaustive comparison of all previously mentioned techniques on seven different noise types and five different SNR values. We also report the WER on clean speech to observe degradation with each noise adaptation technique in place. We make a number of  observations. Among the SE techniques, D\textsc{emucs} which performs best on objective SE scores (as shown in Table~\ref{tab:enhance}), also performs best in the ASR experiments. SE-VCAE suffers from distortion of content in the speech signal, which also reflects in its high degradation of clean speech WER. We observe that DeepXi is outperformed (in most noise conditions, and is otherwise matched) by techniques as simple as DAT. The performance mismatch between the reported numbers in DeepXi~\cite{nicolson2019deep} and our numbers could be attributed to the difference in sizes of our noise datasets; their dataset was much larger in size compared to ours.  Interestingly, even with the relatively smaller amounts of noise samples in our dataset, the D\textsc{emucs} technique is able to generalize well. On the relatively stationary noise types namely, \textit{`Car', `Metro'} and \textit{`Traffic'} (\texttt{Noise A}), D\textsc{emucs} outperforms all techniques including all the model-adaptation techniques. For the relatively non-stationary noise types namely, \textit{`Babble', `Airport/Station',`Cafe'} and \textit{`AC/Vacuum'} (\texttt{Noise B}) D\textsc{emucs} and MTL are statistically very close. AvT yields the largest reductions in WER on \texttt{Noise B} samples (with few exceptions). 

Another important distinction to make between the SE techniques and the ML-based techniques is that the SE techniques rely on high-quality pretrained models as a starting point. With our small noise dataset, training the SE model from scratch would not be an option. In contrast, our ML-based techniques are expressive enough to be able to learn from our limited noise datasets without any prior pretraining. 

The overall takeaways from this investigation are the following: 1) Among the SE techniques, D\textsc{emucs} clearly outperforms the other two SE techniques by large margins. 2) Among the ML-based techniques, AvT is largely the best-performing technique across all noise types, with some exceptions. While adversarial training drives the representations to be noise-invariant and helps the noisy speech WERs (evident in low SNR conditions), it has an adverse effect on clean speech and high SNR conditions. AvT incurs the highest WERs on clean speech WER (after SE-VCAE) and sometimes comes second to MTL in high SNR conditions. 3) MTL and AvT are both significantly better than the DAT techniques. Summarily, either of D\textsc{emucs} or AvT might be a good choice for noise adaptation but the underlying noise type should also factor into the choice. Table~\ref{tab:exp} provides an ablation analysis justifying our choices of LSTM$_3$ and LSTM$_4$ for Soft-Freeze DAT, LSTM$_2$ for MTL and $\lambda_r = 0.05$ for AvT. The other hyperparameters were less influential and were selected based on performance on the development set.

\section{Conclusions}

In this work, we present a detailed comparison of three speech enhancement techniques and three  model-based adaptation techniques for robust E2E ASR across a set of diverse noise types. We observe different trends for different noise types; while adversarial learning yields the largest improvements in performance on non-stationary noise types, a new SE technique D\textsc{emucs} gives the best results on relatively stationary noise types. In future work, we aim to extend our analysis to transformer-based ASR systems and existing noisy datasets.

\label{ssec:results} 
\clearpage
\bibliographystyle{IEEEbib}
\bibliography{ICASSP}

\end{document}